\documentstyle[12pt,epsf,a4]{article}

\newcommand{\beq}{\begin{equation}}
\newcommand{\eeq}{\end{equation}}
\newcommand{\beqa}{\begin{eqnarray}}
\newcommand{\eeqa}{\end{eqnarray}}
\newcommand{\no}{\nonumber}
\newcommand{\q}{\quad}
\newcommand{\qq}{\qquad}
\newcommand{\mnod}{\stackrel{\circ}{M}}

\begin{document}

\hfill 

\hfill 

\bigskip\bigskip

\begin{center}

{{\Large\bf The $\eta'$ in baryon chiral perturbation theory\footnote{Work
supported in part by the BMBF}}}

\end{center}

\vspace{.4in}

\begin{center}
{\large B. Borasoy\footnote{email: borasoy@physik.tu-muenchen.de}}

\bigskip

\bigskip

Physik Department\\
Technische Universit{\"a}t M{\"u}nchen\\
D-85747 Garching, Germany \\

\vspace{.2in}

\end{center}

\vspace{.7in}

\thispagestyle{empty} 

\begin{abstract}
We include in a systematic way the $\eta'$ 
in baryon chiral perturbation theory. 
The most general relativistic 
effective Lagrangian describing the interaction of the lowest lying
baryon octet with the Goldstone boson octet and the $\eta'$ is presented
up to linear order in the derivative expansion and its heavy baryon limit
is obtained.
As explicit examples, we calculate the baryon masses and the $\pi N$
$\sigma$-term up to one-loop order in the heavy baryon formulation.
A systematic expansion in the meson masses is possible, and appearing
divergences are renormalized.
\end{abstract}

\vfill

\section{Introduction}  
Chiral perturbation theory is the effective field theory of QCD
at very low energies. The QCD Lagrangian with massless quarks
exhibits an $SU(3)_R \times SU(3)_L$ chiral symmetry which is broken down
spontaneously to $SU(3)_V$, 
giving rise to a Goldstone boson octet of pseudoscalar mesons
which become massless in the chiral limit of zero quark masses.
On the other hand, the axial $U(1)$ symmetry of the QCD Lagrangian is broken by
the anomaly. 
The corresponding pseudoscalar singlet would otherwise have a mass comparable
to the pion mass \cite{W}. Such a particle is missing in the spectrum and the
lightest candidate would be the $\eta'$ with a mass of 958 MeV which is
considerably heavier than the octet states.
In conventional chiral perturbation theory the $\eta'$ 
is not included explicitly, although it does show up in the form of a
contribution to a coupling coefficient of the Lagrangian, a so-called
low-energy constant (LEC).

Experiment suggests that mixing between the Goldstone boson octet and the
singlet $\eta_0$ occurs. More precisely, the singlet mixes with the uncharged
octet states, $\pi^0$ and $\eta_8$. We will work in the isospin limit of
identical light quark masses, $m_u = m_d$, throughout this work. In this case,
the  $\pi^0$ decouples and only $\eta_0$-$\eta_8$ mixing remains, yielding the
physical states $\eta$ and $\eta'$.
%The pertinent mixing angle can be determined e.g. from the two photon decays 
%of the $\pi^0$, $\eta$ and $\eta'$ leading to an average value of $-20^\circ$:
%the $\eta$ is not a pure Goldstone boson state any longer.

In order to include this effect in chiral perturbation theory one should treat
the $\eta'$ as a dynamical field variable instead of integrating it out from
the effective theory.
This approach is also motivated by large $N_c$ considerations. In this limit
the axial anomaly is suppressed by powers of 1/$N_c$ and gives rise to a ninth 
Goldstone boson, the $\eta'$.

The purpose of this work is to include the $\eta'$ in baryon chiral 
perturbation theory in a systematic fashion without invoking large $N_c$
arguments. It will be shown that observables
can be expanded in the masses of the meson octet and singlet
simultaneously. The relative size of the expansion parameters is given by
$m_\eta / m_{\eta'}$. In this introductory presentation
we will restrict ourselves to
the development of the theory and do not address such 
issues as determination of
the appearing LECs from experiment and convergence of the series.

In the following section, we present the purely mesonic Lagrangian including
the $\eta'$. The extension to the baryonic case is discussed in Section 3. As
explicit examples we present the calculation of the baryon octet masses and the
$\pi N$ $\sigma$-term up to one-loop order. A novel feature of this approach is
the appearance of divergences at one-loop level which are
renormalized in a chiral invariant way.
We conclude with a short summary.

\section{The mesonic Lagrangian}
In this section we will consider the purely mesonic Lagrangian 
including the $\eta'$.
The derivation of this Lagrangian has been given elsewhere, see e.g.
\cite{GL,Leu,H-S}, so we will restrict ourselves to the 
repetition of some of the
basic formulae which are needed in the present work. In \cite{Leu,H-S}
the topological charge operator coupled to an external field is added to the
QCD Lagrangian
\beq  \label{lag}
{\cal L} = {\cal L}_{QCD} - \frac{g^2}{16 \pi^2} \theta(x) \mbox{tr}_c
  ( G_{\mu \nu} \tilde{G}^{\mu \nu} )
\eeq
with $\tilde{G}_{\mu \nu} = \epsilon_{\mu \nu \alpha \beta} G^{\alpha
\beta}$ and $\mbox{tr}_c$ is the trace over the color indices.
Under $U(1)_R \times U(1)_L$ the axial $U(1)$ anomaly 
adds a term $ -( g^2 / 16 \pi^2)
2 N_f \, \alpha \, \mbox{tr}_c ( G_{\mu \nu} \tilde{G}^{\mu \nu} )$ to the
QCD Lagrangian, with $N_f$ being the number of different quark flavors and
$\alpha$ the angle of the global axial $U(1)$ rotation.
The vacuum angle $\theta(x)$ is in this context treated as an external field
that transforms under an axial $U(1)$ rotation as
\beq
\theta(x) \rightarrow  \theta'(x) = \theta(x) - 2 N_f \alpha .
\eeq
Then the term generated by the anomaly in the fermion determinant is
compensated by the shift in the $\theta$ source and the Lagrangian from
Eq.(\ref{lag}) remains invariant under axial $U(1)$ transformations.
The symmetry group $SU(3)_R \times SU(3)_L$ of the Lagrangian ${\cal L}_{QCD}$
is extended to $U(3)_R \times U(3)_L$ for ${\cal L}$.\footnote{To be more
precise, the Lagrangian changes by a total derivative which gives rise to the
Wess-Zumino term. We will neglect this contribution since the corresponding
terms involve five or more meson fields which do not play any role for the
discussions here.}
This property remains at the level of an effective theory and the
additional source $\theta$ also shows up in the effective Lagrangian.
Let us consider the purely mesonic effective theory first.
The lowest lying pseudoscalar meson nonet is summarized in a matrix valued 
field $U(x)$
\begin{equation}
 U(\phi,\eta_0) = u^2 (\phi,\eta_0) = 
\exp \lbrace 2 i \phi / F_\pi + i \sqrt{\frac{2}{3}} \eta_0/ F_0 \rbrace  
, 
\end{equation}
where $F_\pi \simeq 92.4$ MeV is the pion decay constant and the singlet
$\eta_0$ couples to the singlet axial current with strength $F_0$.
The unimodular part of the field $U(x)$ contains the degrees of freedom of
the Goldstone boson octet $\phi$
\begin{eqnarray}
 \phi =  \frac{1}{\sqrt{2}}  \left(
\matrix { {1\over \sqrt 2} \pi^0 + {1 \over \sqrt 6} \eta_8
&\pi^+ &K^+ \nonumber \\
\pi^-
        & -{1\over \sqrt 2} \pi^0 + {1 \over \sqrt 6} \eta_8 & K^0
        \nonumber \\
K^-
        &  \bar{K^0}&- {2 \over \sqrt 6} \eta_8  \nonumber \\} 
\!\!\!\!\!\!\!\!\!\!\!\!\!\!\! \right) \, \, \, \, \, ,  
\end{eqnarray}
while the phase det$U(x)=e^{i\sqrt{6}\eta_0/F_0}$
describes the $\eta_0$ .
The symmetry $U(3)_R \times U(3)_L$ does not have a dimension-nine
irreducible representation and consequently does not exhibit a nonet symmetry. 
We have therefore used the different notation $F_0$ for the decay constant 
of the singlet field.
The effective Lagrangian is formed with the fields $U(x)$,  
derivatives thereof and also includes both the quark mass matrix ${\cal M}$ and
the vacuum angle $\theta$: ${\cal L}_{\mbox{eff}}(U,\partial U,\ldots,
{\cal M},\theta)$. Under $U(3)_R \times U(3)_L$ the fields transform as
follows
\beq
U' = RUL^{\dagger} \q , \qq {\cal M}'= R{\cal M}L^{\dagger} \q , \qq
\theta'(x) = \theta(x) - 2 N_f \alpha
\eeq
with $R \in U(3)_R$, $L \in U(3)_L$,
but the Lagrangian remains invariant. 
In order to incorporate the baryons into the effective theory it is convenient
to form an object of axial-vector type with one derivative
\beq
u_{\mu} = i u^\dagger \nabla_{\mu} U u^\dagger
\eeq
with $\nabla_{\mu}$ being the covariant derivative of $U$.
The matrix $ u_{\mu}$ transforms under $U(3)_R \times U(3)_L$
as a matter field,
\beq
u_{\mu} \rightarrow u_{\mu}' = K u_{\mu} K^\dagger
\eeq
with $K(U,R,L)$ the compensator field representing an element of the conserved
subgroup $ U(3)_V$.
The phase of the determinant
det$U(x)=e^{i\sqrt{6}\eta_0/F_0}$ transforms  under axial $U(1)$ as
$\sqrt{6} \eta_0'/F_0 = \sqrt{6} \eta_0/F_0  + 2 N_f \alpha$ so that
the combination $\sqrt{6} \eta_0/F_0 + \theta$ remains invariant.
It is more convenient to replace the variable $\theta$ by this invariant
combination, ${\cal L}_{\mbox{eff}}= {\cal L}_{\mbox{eff}}(U,\partial U,\ldots,
{\cal M},\sqrt{6} \eta_0/F_0 + \theta)$.
One can now construct the effective Lagrangian in these
fields that respects the symmetries of the underlying theory.
In particular, the Lagrangian is invariant under $U(3)_R \times U(3)_L$
rotations of $U$ and ${\cal M}$ at a fixed value of the last argument.
The most general Lagrangian up to and including terms with two derivatives and
one factor of ${\cal M}$ reads
\beq  \label{mes}
{\cal L}_{\phi} = - V_0 + V_1 \langle u_{\mu} u^{\mu} \rangle 
+ V_2 \langle \chi_+ \rangle + i V_3 \langle \chi_- \rangle
+ V_4 \langle u_{\mu} \rangle \langle u^{\mu} \rangle .
\eeq
The expression $\langle \ldots \rangle$ denotes the trace in flavor space
and the quark mass matrix ${\cal M} = \mbox{diag}(m_u,m_d,m_s)$
enters in the combinations
\beq
\chi_\pm = 2 B_0 ( u {\cal M} u \pm  u^\dagger {\cal M} u^\dagger)
\eeq
with $B_0 = - \langle  0 | \bar{q} q | 0\rangle/ F_\pi^2$ the order
parameter of the spontaneous symmetry violation.
Note that a term of the type $\langle u_{\mu} \rangle \nabla^{\mu} \theta$
can be transformed away \cite{Leu} 
and a term proportional to $\nabla_{\mu} \theta
\nabla^{\mu} \theta$ does not enter the calculations performed in the present
work and will be neglected.
The coefficients $V_i$ are functions of the variable 
$\sqrt{6} \eta_0/F_0 + \theta$, $V_i(\sqrt{6} \eta_0/F_0 + \theta)$,
and can be expanded in terms of this variable. At a given order of
derivatives of the meson fields $U$ and insertions of the quark mass matrix 
${\cal M}$ one obtains an infinite string of increasing powers of the 
singlet field $\eta_0$ with couplings which are not fixed by chiral symmetry.
The terms $V_{1,\ldots,4}$ are of second chiral order, whereas $V_0$
is of zeroth chiral order.
Parity conservation implies that the $V_i$ are all even functions
of $\sqrt{6} \eta_0/F_0 + \theta$ except $V_3$, which is odd, and
$V_1(0) = V_2(0) = F_\pi^2/4$ gives the correct  normalizaton
for the quadratic terms of the Goldstone boson octet.
The kinetic energy of the the $\eta_0$ singlet field obtains contributions
from $V_1 \langle u_{\mu} u^{\mu} \rangle $ and
$V_4 \langle u_{\mu} \rangle \langle u^{\mu} \rangle$ which read
\beq
\Big( \frac{F_\pi^2}{2 F_0^2} + \frac{6}{F_0^2} V_4(0) \Big) 
\partial_\mu \eta_0 \partial^\mu \eta_0  .
\eeq
We renormalize the $\eta_0$ field in such a way that the coefficient in
brackets is 1/2 in analogy to the kinetic term of the octet.
By redefining $F_0$ and keeping for simplicity the same notation both
for $\eta_0$ and $F_0$ one arrives at the same Lagrangian as in 
Eq.(\ref{mes}) but with  $V_4(0) = (F_0^2-F_\pi^2)/12$ in order to ensure
the usual normalization for the kinetic term of a pseudoscalar particle.

For our considerations here we can safely neglect the source $\theta$.
The coefficients $V_i$ are then functions of $\eta_0$ only, $V_i(\eta_0)$,
and their Taylor expansions read at lowest orders, e.g.,
\beqa  \label{para}
V_0 &=& \mbox{const.} + v \, \eta_0^2 + \ldots \no \\
V_2 &=& \frac{1}{4} F_\pi^2 + w \, \eta_0^2 + \ldots \no \\
V_3 &=& x \, \eta_0 + \ldots
\eeqa
where the ellipses denote terms with higher powers of $\eta_0$.
Here, we presented only the terms which enter our calculation.
Note that the quadratic term in $V_0$ contributes to the $\eta_0$ mass
which does not vanish in the chiral limit, i.e. the 
$\eta_0$ is not a Goldstone boson.

Expanding the Lagrangian in terms of the meson fields
one observes terms quadratic in the meson fields that contain the factor
$\eta_0 \eta_8$ which leads to $\eta_0$-$\eta_8$ mixing.
Such terms arise from the explicitly symmetry breaking terms 
$V_2 \langle \chi_+ \rangle + i V_3 \langle \chi_- \rangle$ and read
\beq
- \Big( \frac{2 \sqrt{2}}{3} \frac{F_\pi}{F_0} + \frac{8}{\sqrt{3}} 
\frac{1}{F_\pi} x \Big) B_0 (\hat{m} -m_s) \eta_0 \eta_8
\eeq
with $\hat{m} = \frac{1}{2} (m_u + m_d)$.
The states $\eta_0$ and $\eta_8$ are therefore not mass eigenstates.
The mixing yields the eigenstates $\eta$ and $\eta'$,
\beqa
| \eta \rangle & = & \cos \theta \, | \eta_8 \rangle -
                     \sin \theta \, | \eta_0 \rangle  \no \\
| \eta'\rangle & = & \sin \theta \, | \eta_8 \rangle +
                     \cos \theta \, | \eta_0 \rangle , 
\eeqa
where we have neglected other pseudoscalar isoscalar states which could mix
with both $\eta_0$ and $\eta_8$ and we assume that the mixing parameters do
not depend on the energy of the state.
The $\eta$-$\eta'$ mixing angle can be determined from the two
photon decays of $\pi^0, \eta, \eta'$, which require a mixing angle
around -20$^\circ$ \cite{exp}. 
We will make use of this experimental input in order to
diagonalize the mass terms of the effective mesonic Lagrangian. Since we work
in the isospin limit $m_u=m_d$, this is the only mixing between the meson
states.

\section{Inclusion of baryons}
We now proceed by including the lowest lying baryon octet into the effective
theory. The baryon octet $B$ is given by the matrix
\beqa
 B =  \left(
\matrix { {1\over \sqrt 2} \Sigma^0 + {1 \over \sqrt 6} \Lambda
&\Sigma^+ &p \nonumber \\
\Sigma^-
        & -{1\over \sqrt 2} \Sigma^0 + {1 \over \sqrt 6} \Lambda & n
        \nonumber \\
\Xi^-
        &  \Xi^0 &- {2 \over \sqrt 6} \Lambda  \nonumber \\} 
\!\!\!\!\!\!\!\!\!\!\!\!\!\!\! \right) \, \, \, \, \,  
\end{eqnarray}
which transforms as a matter field
\beq
B \rightarrow B' = K B K^\dagger .
\eeq
Up to linear order in the derivative expansion the most general relativistic 
effective Lagrangian describing the interaction of the baryon octet with 
the meson nonet reads
\beqa  \label{bar0}
{\cal L}_{\phi B} &=& i W_1 \langle [D^{\mu},\bar{B}]\gamma_{\mu} B
\rangle - i W_1^* \langle \bar{B}  \gamma_{\mu}  [D^{\mu},B] \rangle 
+ W_2 \langle \bar{B}B \rangle \no \\
&& + W_3 \langle \bar{B} \gamma_{\mu}
 \gamma_5 \{u^{\mu},B\} \rangle  
+  W_4 \langle \bar{B} \gamma_{\mu} \gamma_5 [u^{\mu},B] \rangle 
+ W_5 \langle \bar{B} \gamma_{\mu} \gamma_5 B \rangle \langle u^{\mu} \rangle
\no \\
&& + W_6 \langle \bar{B} \gamma_{\mu} \gamma_5 B \rangle \nabla^{\mu} \theta
+ i W_7 \langle \bar{B} \gamma_5 B \rangle 
\eeqa
with $D_{\mu}$ being the covariant derivative of the baryon fields.
The $W_i$ are functions of the combination $\sqrt{6} \eta_0/F_0 + \theta$.
From parity it follows that they are even in  this variable except
$W_7$ which is odd.
One can further reduce the number of independent terms by making the
following transformation.
By decomposing the baryon fields into their left- and right handed 
components
\beq
B_{R/L} = \frac{1}{2} ( 1 \pm \gamma_5) B
\eeq
and transforming the left- and right-handed states separately via
\beqa
B_{R/L} &\rightarrow& \frac{1}{\sqrt{W_2 \pm i W_7}} B_{R/L} \no \\
\bar{B}_{R/L} &\rightarrow& \frac{1}{\sqrt{W_2 \mp i W_7}} \bar{B}_{R/L} 
\eeqa
one can eliminate the $\langle \bar{B} \gamma_5 B \rangle $ term
and simplify the coefficient of $\langle \bar{B}B \rangle$. The details of this
calculation are given in App. A.
The Lagrangian in Eq.(\ref{bar0}) reduces then to
\beqa  \label{bar}
{\cal L}_{\phi B} &=& i U_1 \Big( \langle [D^{\mu},\bar{B}]\gamma_{\mu} B
\rangle - \langle \bar{B}  \gamma_{\mu}  [D^{\mu},B] \rangle \Big)
- \mnod \langle \bar{B}B \rangle + U_2 \langle \bar{B} \gamma_{\mu}
 \gamma_5 \{u^{\mu},B\} \rangle  \no \\
&& +  U_3 \langle \bar{B} \gamma_{\mu} \gamma_5 [u^{\mu},B] \rangle 
+ U_4 \langle \bar{B} \gamma_{\mu} \gamma_5 B \rangle \langle u^{\mu} \rangle
+ U_5 \langle \bar{B} \gamma_{\mu} \gamma_5 B \rangle \nabla^{\mu} \theta 
\eeqa
with $\mnod$ being the baryon octet mass in the chiral limit.
The coefficients $U_i$ are real and even functions of
$\sqrt{6} \eta_0/F_0 + \theta$. The last term which includes the derivative of
the source $\theta$ can be disregarded for the processes considered here,
and we can set $\theta=0$ for our purposes. The expansion of the
coefficients $U_i$ in terms of $\eta_0$ read
\beqa
U_1 &=& - \frac{1}{2} + \lambda_1 \, \eta_0^2 + \ldots \no \\
U_2 &=& - \frac{1}{2} D + \ldots \no \\
U_3 &=& - \frac{1}{2} F + \ldots \no \\
U_4 &=& \lambda_2 + \ldots 
\eeqa
where the ellipses denote higher orders in $\eta_0$ and we have only shown
terms that contribute to the baryon masses up to one-loop order.
The axial-vector couplings $D$ and $F$ can be determined from semileptonic
hyperon decays. A fit to the experimental data
%without the inclusion of the $\eta'$ 
delivers 
$D=0.80 \pm 0.01$ and $F=0.46 \pm 0.01$ \cite{CR}.

The drawback of the relativistic framework including baryons is that 
due to the existence of a new mass scale, namely the baryon mass
in the chiral limit $\mnod$, there exists no strict
chiral counting scheme, i.e. a one-to-one correspondence between the meson
loops and the chiral expansion. In order to overcome this problem
%in conventional baryon chiral perturbation theory
one integrates out the heavy degrees of freedom of the baryons, similar to a
Foldy-Wouthuysen transformation, so that a chiral counting scheme emerges.
We will adopt this procedure here 
in order to write down the heavy baryon formulation of the theory
in the presence of the singlet field.
Observables can then be expanded simultaneously in the Goldstone boson 
octet masses and the $\eta'$ mass that does not vanish in the chiral limit.
One obtains a one-to-one correspondence between the meson loops and
the expansion in their masses and derivatives both for octet and singlet.
Thus, even in the presence of the massive $\eta'$ field a strict chiral
counting scheme is possible. The purpose of this presentation is to show 
that such a scheme can be established. Issues as the convergence of the
expansion in the $\eta'$ mass which is in size comparable to the baryon octet
masses  will not be addressed here.
Here, we only mention that the large $N_c$ scheme provides some motivation why
higher orders in the $\eta'$ mass could be suppressed, since
such terms arise from
higher orders in $\eta_0$ in the expansions of the coefficients $U_i$ and $V_i$
of the Lagrangian and are suppressed by powers of $1/N_c$.
By neglecting the contributions from the singlet and treating $\eta_8$ as
$\eta$ one would again arrive at the conventional chiral expansion.
The relative size of the expansion parameters is 
given by $m_\eta/m_{\eta'}$, which is of order $\sqrt{N_c m_s}$ with $N_c$ the
number of colors and $m_s \gg \hat{m}$ has been assumed.

The process of integrating out the heavy degrees of freedom 
of the baryons from the effective
theory is rather well known \cite{Man} and we only present the result here.
A four-velocity $v$ is assigned to the baryons and
the heavy baryon Lagrangian reads to the order we are working
\beqa  \label{hb}
{\cal L}_{\phi B} &=& i  \langle \bar{B}  [v \cdot D,B] \rangle
+ \lambda_1 \eta_0^2 \Big( - 2 \mnod \langle \bar{B}B \rangle 
+ i \langle [v \cdot D,\bar{B}] B \rangle
- i \langle \bar{B}  [v \cdot D,B] \rangle \Big) \no \\
&& - D \langle \bar{B} S_{\mu} \{u^{\mu},B\} \rangle 
 - F \langle \bar{B} S_{\mu} [u^{\mu},B] \rangle 
+ 2 \lambda_2 \langle \bar{B} S_{\mu}  B \rangle \langle u^{\mu} \rangle
\eeqa
where we omitted higher powers in the singlet field $\eta_0$ since they do
not contribute to the lowest non-analytic contributions for the baryon masses
and the ${\pi N}$ $\sigma$-term.
The Dirac algebra simplifies considerably and $2 S_\mu = i \gamma_5 \sigma_{\mu
\nu} v^\nu$ denotes the Pauli-Lubanski spin vector.

\section{Baryon masses}
In this section we present the calculation of the baryon octet masses within
the new framework of heavy baryon chiral perturbation theory including the
$\eta'$. In this work our main concern is to include the $\eta'$ in a
systematic way and, therefore, we restrict ourselves to the calculation of the
one-loop diagrams of the $\eta$ and $\eta'$ 
with the Lagrangian given in Eq.(\ref{hb}). A complete
analysis of the baryon masses would also require the inclusion of explicitly
symmetry breaking terms and the $\pi , K$ loops  \cite{BKM,BM}.
We will disregard such terms and consider only the 
lowest non-analytic contributions of the $\eta$ and $\eta'$ to the 
baryon masses. 
The contributing one-loop diagrams of the Lagrangian in Eq.(\ref{hb}) 
are depicted in Figures 1 and 2. 
In addition to the usual self-energy diagram, Fig.~1, that
already enters the calculation in $SU(3)$ chiral perturbation theory, one
obtains a tadpole diagram with a singlet loop, Fig.~2. 
This singlet diagram delivers a
constant contribution to all baryon masses and is therefore not observable
since it can be absorbed in a redefinition of $\mnod$.
The contributions of the $\eta$ and $\eta'$ fields from Figs. 1 and 2 
are given by
\beqa \label{masses}
\delta M_B &=& - \frac{1}{24 \pi F_\pi^2} \alpha_B^2 \, \Big(
           m_\eta^3 \cos^2 \theta + m_{\eta'}^3 \sin^2 \theta  \Big)  \no \\
           && - \frac{\sqrt{2}}{24 \pi F_0 F_\pi} [D-3 \lambda_2] \alpha_B \,
           \sin 2 \theta   \, \Big(m_{\eta'}^3 - m_\eta^3 \Big) \no \\
           && - \frac{1}{12 \pi F_0^2} [D-3 \lambda_2]^2 \, \Big(
           m_\eta^3 \sin^2  \theta + m_{\eta'}^3 \cos^2 \theta  \Big) \no \\
           && + \frac{1}{8 \pi^2} \mnod \lambda_1  \Big( \cos^2 \theta \, 
           m_{\eta'}^2 \ln \frac{m_{\eta'}^2}{\mu^2} + \sin^2  \theta\, 
           m_\eta^2 \ln \frac{m_\eta^2}{\mu^2}
           \Big) + \Delta  
\eeqa
with $\mu$ being the scale introduced in dimesional regularization and
a divergent piece from the tadpole
\beq
\Delta = 4 \mnod \lambda_1 \Big( m_{\eta'}^2 \cos^2 \theta + 
         m_\eta^2 \sin^2  \theta    \Big) L,
\eeq
with
\beq
L = \frac{\mu^{d-4}}{16 \pi^2} \Big\{ \frac{1}{d-4} - \frac{1}{2} 
   [ \ln 4 \pi + 1 - \gamma_E] \Big\}.
\eeq
Here, $\gamma_E = 0.5772215$ is the Euler-Mascheroni constant.
The coefficients $\alpha_B$ are given by
\beqa
\alpha_N &=& \frac{1}{2} (D-3F) , \qq  \qq \alpha_\Lambda = -D ,
\qq  \qq \alpha_\Sigma =D   , \no \\
\alpha_\Xi &=& \frac{1}{2} (D+3F).
\eeqa
The contributions of the other mesons remain unchanged after the inclusion of
the singlet field and are given elsewhere, see e.g. \cite{BKM}.
We have expressed our results in terms of the physical states
$\eta$ and $\eta'$. The crucial difference with respect
to heavy baryon chiral perturbation
theory without the singlet is the appearance of the tadpole contribution which
delivers  terms quadratic in the meson masses $m_\eta$ and $m_{\eta'}$. This
seems to violate the observation that the lowest non-analytic pieces start
contributing at third chiral order \cite{GZ}, 
but both the $\eta$ and the $\eta'$ masses
contain a piece from the singlet field which does not vanish in the chiral
limit. 
The $\eta'$ mass and the logarithm thereof can be expanded around this value 
which leads to a series analytic in the quark masses,
whereas for the $\eta$ we note that the mixing angle is of chiral order
${\cal O}(p^2)$ for small quark masses.
Therefore, Eq.(\ref{masses}) does not contradict the results from \cite{GZ}.

The tadpole is divergent and has to be renormalized. This is a new feature
at leading order
which did not occur in conventional heavy baryon chiral perturbation
theory. Its divergent piece can be renormalized by redefining both the
baryon mass in the chiral limit $\mnod$ and the coefficient $b_0$ 
of the $SU(3)$ 
invariant piece of the explicitly symmetry breaking terms, $\langle \bar{B} B
\rangle \langle \chi_+ \rangle$. 
To this end, one uses the relation between the masses of the physical states
$\eta$ and $\eta'$ and the coefficient of the mass term 
for $\eta_0$ in ${\cal L}_\phi$: 
\beq
m_{\eta'}^2 \cos^2 \theta +   m_\eta^2 \sin^2  \theta  =
\Big(\frac{2 F_\pi^2}{3 F_0^2}
          - 8w + 8 \sqrt{\frac{2}{3}} \frac{x}{F_0} \Big)\, B_0 \,
           ( 2 \hat{m} + m_s) + 2 v 
\eeq
with $w,v$ and $x$ being parameters of the mesonic Lagrangian as given in
Eq.(\ref{para}). 
The last term  without the quark masses is absorbed by $\mnod$, whereas a
redefinition of $b_0$ renormalizes the quark mass dependent divergences.
\beqa
\mnod   &\rightarrow&  \stackrel{\circ}{M^r} - 8 \mnod \lambda_1 v L   \\
b_0    &\rightarrow&  b_0^r + \mnod \lambda_1  \Big(\frac{2 F_\pi^2}{3 F_0^2}
          - 8w + 8 \sqrt{\frac{2}{3}} \frac{x}{F_0} \Big) L .  \label{b0}
\eeqa

But before proceeding, it is worthwhile taking a closer look at the
contributions of the $\eta'$ to $M_\Lambda$ and $M_\Sigma$. 
The $SU(3)$-breaking contributions of the $\eta'$ read for these two cases
\beq  \label{lam}
 - \frac{1}{24 \pi F_\pi^2} D^2 \, m_{\eta'}^3 \sin^2 \theta 
         \pm \frac{\sqrt{2}}{24 \pi F_0 F_\pi} [D-3 \lambda_2] D\,
           \sin 2 \theta  \, m_{\eta'}^3 .
\eeq
Inserting the values $m_{\eta'} = 958$ MeV and $\theta = -20^\circ$ and using
the central
values for $D$ and $F$ from \cite{CR}\footnote{We assume here that the
inclusion of the $\eta'$ does not alter the values for $D$ and $F$
significantly.}, both
contributions are relatively small only if $D \simeq 3 \lambda_2$
yielding a mass shift of $-103$ MeV in both cases. For all
other values of $\lambda_2$ one obtains substantial $SU(3)$ breaking
contributions of the $ \eta'$ to the baryon masses. In order to get a rough 
estimate we set $\lambda_2$ equal to zero
and use $F_0= F_\pi = 92.4 $ MeV.  From Eq.(\ref{lam}) one obtains 
the numerical values $-907$ MeV and $701$ MeV for the
$\Lambda$ and $\Sigma$ mass shifts, respectively.
A more quantitative statement about these 
contributions can only be made if one has a
reliable estimate of the parameter $\lambda_2$, but this is beyond the scope of
this work.

\section{The $\pi N $ $\sigma$-term}
Closely related to the nucleon mass is the $\pi N $ $\sigma$-term
\beq
\sigma_{\pi N}(t) = \hat{m} \langle p'| \bar{u}u + \bar{d}d | p \rangle
\eeq
with $| p \rangle$ a proton state with momentum $p$ and $t= (p'-p)^2$ the
momentum transfer squared. The $\sigma$-term vanishes in the chiral limit of
zero quark masses and measures the scalar quark density inside the proton. 
Thus it
is particularly suited to test our understanding of spontaneous and explicit
chiral symmetry breaking.
At zero momentum transfer squared, $t=0$, the $\pi N $ $\sigma$-term is related
to the nucleon mass $M_N$ via the Feynman-Hellmann theorem
\beq
\sigma_{\pi N}(0) = \hat{m} \frac{\partial M_N}{\partial \hat{m}} .
\eeq
Again we will restrict ourselves to the presentation of $\eta$ and $\eta'$
loops. Contributions from the contact terms of second chiral order and $\pi ,
K$ loops have already been calculated in \cite{BKM}.
The result is
\beqa
\sigma_{\pi N}(0) &=& - \frac{1}{64 \pi F_\pi^2} [D-3F]^2 \, m_\pi^2 \, \Big(
               {\cal A} \sin^2 \theta \, m_{\eta'} + {\cal B} \cos^2  \theta \,
               m_\eta  \Big)  \no \\
&& - \frac{1}{8 \pi F_0^2} [D-3 \lambda_2]^2 \, m_\pi^2 \, \Big(
              {\cal A} \cos^2 \theta \, m_{\eta'} + {\cal B} \sin^2 \theta \,
                 m_\eta  \Big)  \no \\
&&  + \frac{\sqrt{2}}{32 \pi F_0 F_\pi} [D-3 \lambda_2] [D-3F]\,  m_\pi^2 \, 
      \sin 2 \theta \, \Big( {\cal A} m_{\eta'} - {\cal B} m_\eta \Big) \no \\
&& + \frac{1}{8 \pi^2} \mnod \lambda_1 m_\pi^2 \, \Big( 
      {\cal A}  \cos^2 \theta \,\, [1+\ln \frac{m_{\eta'}^2}{\mu^2}]
   + {\cal B} \sin^2 \theta \, \, [1+\ln \frac{m_\eta^2}{\mu^2}]\Big)
  + \Delta' \no \\
\eeqa
with the coefficients
\beqa
{\cal A}  & = &  \frac{1}{3}  \sin^2 \theta + \Big( \frac{2 F_\pi^2}{3 F_0^2}
          - 8w + 8 \sqrt{\frac{2}{3}} \frac{x}{F_0} \Big) \cos^2 \theta \no \\
&&  + \Big( \frac{\sqrt{2} F_\pi}{3 F_0} + 4 \frac{x}{\sqrt{3}F_\pi} \Big)
    \sin 2 \theta  \\ \no \\
{\cal B}  & = &   \frac{1}{3}  \cos^2 \theta + \Big( \frac{2 F_\pi^2}{3 F_0^2}
          - 8w + 8 \sqrt{\frac{2}{3}} \frac{x}{F_0} \Big) \sin^2 \theta \no \\
&& - \Big( \frac{\sqrt{2} F_\pi}{3 F_0} + 4 \frac{x}{\sqrt{3}F_\pi} \Big)
    \sin 2 \theta   
\eeqa
and a divergent piece
\beq
\Delta' =  4 \mnod \lambda_1 m_\pi^2  \, \Big( {\cal A}  \cos^2 \theta +
           {\cal B} \sin^2 \theta \Big) \, L .
\eeq
This time the divergent part is renormalized by the term
$\langle \bar{B} B  \rangle \langle \chi_+ \rangle$ only, since $\mnod$ does
not contribute to the $\sigma$-term. 
Using the same renormalization prescription for $b_0$ as in Eq.(\ref{b0}) one
achieves cancellation of the divergences.

Of particular interest is the shift  of the $\sigma$-term to the Cheng-Dashen
point $t= 2 m_\pi^2$. It is convenient to work in the Breit frame, $v \cdot p =
v \cdot p'$. The  $\eta$ and $\eta'$ contributions to the
$\sigma$-term shift read
\beqa
&& \sigma_{\pi N}(2 m_\pi^2) - \sigma_{\pi N}(0)  = \no \\
&& - \frac{1}{8 \pi} \, m_\pi^2 \,
   \Big( \frac{1}{24 F_\pi^2} [D-3F]^2  \sin^2 \theta +
       \frac{1}{3 F_0^2} [D-3 \lambda_2]^2 \cos^2 \theta \no \\
&& - \frac{\sqrt{2}}{12 F_0 F_\pi} [D-3 \lambda_2] [D-3F]\sin 2 \theta \,\Big)
\, {\cal A} \, \Big( - m_{\eta'} + \frac{ m_{\eta'}^2 - m_\pi^2}{
   \sqrt{2} m_\pi} \ln \frac{ \sqrt{2} m_{\eta'} + m_\pi}{\sqrt{2} m_{\eta'} -
    m_\pi} \Big) \no \\
&& - \frac{1}{8 \pi} \, m_\pi^2 \,
   \Big( \frac{1}{24 F_\pi^2} [D-3F]^2  \cos^2 \theta +
       \frac{1}{3 F_0^2} [D-3 \lambda_2]^2 \sin^2 \theta \no \\
&& + \frac{\sqrt{2}}{12 F_0 F_\pi} [D-3 \lambda_2] [D-3F]\sin 2 \theta \,\Big)
\, {\cal B} \, \Big( - m_\eta  + \frac{ m_\eta^2 - m_\pi^2}{
   \sqrt{2} m_\pi} \ln \frac{ \sqrt{2} m_\eta + m_\pi}{\sqrt{2} m_\eta -
    m_\pi} \Big) \no \\
&& + \frac{1}{4 \pi^2} \mnod \lambda_1 m_\pi^2 \, \Big( 
      {\cal A}  \cos^2 \theta \,\, 
      \Big[ - 1+ \sqrt{ \frac{ 2 m_{\eta'}^2}{m_\pi^2}
      - 1} \; \arcsin \frac{ m_\pi}{ \sqrt{2} m_{\eta'}} \Big]\no \\
&& \qq \qq \qq \q + {\cal B} \sin^2 \theta \, \, 
      \Big[ - 1+ \sqrt{ \frac{ 2 m_\eta^2}{m_\pi^2}
      - 1} \; \arcsin \frac{ m_\pi}{ \sqrt{2} m_\eta} \Big]  \Big) .  
\eeqa
In order to give some estimate on the numerical size of the new terms involving
the $\eta'$ as compared to conventional $SU(3) \times SU(3)$ chiral
perturbation theory, we consider the two cases $\lambda_2 = 0$ and $\lambda_2 =
D/3$ with vanishing $w,x,\lambda_1$.
One obtains $0.21$ MeV and $0.02$ MeV for the first and second case, 
respectively.
The general expression of the $\eta$ and $\eta'$ contributions to 
the $\sigma$-term shift reads, in units of MeV,
\beqa 
\sigma_{\pi N}(2 m_\pi^2) - \sigma_{\pi N}(0)  & = & 
0.2 - 1.8 w + 15.3 x + \lambda_2 ( -1.2 + 14.1 w - 106.9 x) \no \\
&& + \lambda_2^2 ( 1.9 -28.1 w + 199.3 x) .
\eeqa
The tadpole contribution proportional to $\lambda_1$
is for realistic values of the parameters about
$10^{-3}$ MeV and has been neglected here.
On the other hand, 
the $\sigma$-term shift depends significantly on the other 
parameters, particularly $x$. It would be desirable to have a reliable estimate
on these parameters in order to make a more quantitative statement about the
$\eta'$ contributions to the $\sigma$-term shift.
These results can be compared with the $\eta$ contribution of about $0.01$ MeV 
from a calculation without the
$\eta'$ in conventional chiral perturbation theory. Note, however, that this
number is only a small fraction of a calculation including both pion
and kaon loops. The $\sigma$-term shift to the Cheng-Dashen point is dominated
by the pion loop contribution and is at one-loop order about $7.5$ MeV
\cite{BKM}.
The $\sigma$-term for general $t$ is given in App.~B.

\section{Summary}
In this work we included in a systematic way the $\eta'$ in baryon chiral
perturbation theory. After setting up the most general relativistic Lagrangian
to first order in the derivative expansion we derived its heavy baryon
limit. In the heavy baryon formulation a one-to-one correspondence between the
number of octet and singlet meson loops and the expansion in the pertinent
masses emerges. The relative size of the expansion parameters is given by
$m_\eta / m_{\eta'}$.

As explicit examples we presented the calculation of the baryon masses and the
$\pi N$ $\sigma$-term up to one loop order in this new framework. In the case
of the baryon masses it turns out that there are sizeable contributions of the
$\eta'$, unless a certain combination of LECs not fixed by chiral
symmetry  happens to be small.

A novel feature of this approach is the appearance of a tadpole diagram at
leading order in the expansion which delivers a divergence. The divergent piece
can be compensated by redefining some of the low-energy constants.
This work introduces the $\eta'$ as a dynamical field variable without invoking
large $N_c$ arguments. Other issues such
as the convergence of the expansion in the
meson masses or the connection to large $N_c$ baryon chiral
perturbation theory will be addressed in future work.

\section*{Acknowledgments}
The author wishes to thank S. Bass,
N. Kaiser, and W. Weise for useful discussions and suggestions.

\appendix 
\def\theequation{\Alph{section}.\arabic{equation}}
\setcounter{equation}{0}
\section{} \label{app:a}
In this appendix, we present the calculation which reduces the Lagrangian
of Eq.(\ref{bar0}) to the one given in Eq.(\ref{bar}).
The starting point is the relativistic Lagrangian
\beqa  
{\cal L}_{\phi B} &=& i W_1  \langle [D^{\mu},\bar{B}]\gamma_{\mu} B
\rangle - i W_1^* \langle \bar{B}  \gamma_{\mu}  [D^{\mu},B] \rangle 
+ W_2 \langle \bar{B}B \rangle \no \\
&& + W_3 \langle \bar{B} \gamma_{\mu}
 \gamma_5 \{u^{\mu},B\} \rangle  
+  W_4 \langle \bar{B} \gamma_{\mu} \gamma_5 [u^{\mu},B] \rangle 
+ W_5 \langle \bar{B} \gamma_{\mu} \gamma_5 B \rangle \langle u^{\mu} \rangle
\no \\
&& + W_6 \langle \bar{B} \gamma_{\mu} \gamma_5 B \rangle \nabla^{\mu} \theta
+ i W_7 \langle \bar{B} \gamma_5 B \rangle .
\eeqa
By decomposing the baryon fields into their left- and right-handed 
components
\beq
B_{R/L} = \frac{1}{2} ( 1 \pm \gamma_5) B
\eeq
and transforming the left- and right-handed states separately via
\beqa
B_{R/L} &\rightarrow& \frac{1}{\sqrt{W_2 \pm i W_7}} B_{R/L} \no \\
\bar{B}_{R/L} &\rightarrow& \frac{1}{\sqrt{W_2 \mp i W_7}} \bar{B}_{R/L} 
\eeqa
the single terms of the Lagrangian transform as follows:
\beq
W_2 \langle \bar{B}B \rangle  + i W_7 \langle \bar{B} \gamma_5 B \rangle
        \q  \rightarrow  \q  \langle \bar{B}B \rangle ;
\eeq
\beqa  \label{tra}
&&  i W_1  \langle [D^{\mu},\bar{B}]\gamma_{\mu} B
\rangle - i W_1^* \langle \bar{B}  \gamma_{\mu}  [D^{\mu},B] \rangle 
    \q \rightarrow \no \\
&&  \qq \qq \qq    \frac{i}{2} \frac{W_1 + W_1^* }{\sqrt{W_2^2 + W_7^2}}
\Big( \langle [D^{\mu},\bar{B}]\gamma_{\mu} B
\rangle - \langle \bar{B}  \gamma_{\mu}  [D^{\mu},B] \rangle \Big) \no \\
&& \qq \qq \qq  
    + \frac{W_1 + W_1^*}{2 \,[W_2^2 + W_7^2]^{3/2}} ( W_7 \partial_\mu W_2 -
    W_2 \partial_\mu W_7 ) \langle \bar{B} \gamma_{\mu} \gamma_5 B \rangle ,
\eeqa
whereas the terms $W_3$ to $W_6$ remain invariant under this transformation.
The last term in Eq.(\ref{tra}) can be absorbed into $W_5$ and $W_6$.
We rescale the baryon fields in such a way that at lowest order the coefficient
of the kinetic term is $-1/2$. From matching to conventional 
$SU(3) \times SU(3)$ baryon chiral perturbation theory it follows that the
coefficient of the mass term $\langle \bar{B}B \rangle$ is $-\mnod$ and
we arrive at the Lagrangian given in Eq.(\ref{bar}).

\setcounter{equation}{0}
\section{} \label{app:b}
For general $t$ the contributions of the $\eta$ and $\eta'$ to the
$\pi N$ $\sigma$-term are given by
\beqa
&& \sigma_{\pi N}(t)  = \no \\
&& - \frac{1}{8 \pi} \, m_\pi^2 \,
   \Big( \frac{1}{24 F_\pi^2} [D-3F]^2  \sin^2 \theta +
       \frac{1}{3 F_0^2} [D-3 \lambda_2]^2 \cos^2 \theta \no \\
&& - \frac{\sqrt{2}}{12 F_0 F_\pi} [D-3 \lambda_2] [D-3F]\sin 2 \theta \,\Big)
\, {\cal A} \, \Big( 2 m_{\eta'} + \frac{ m_{\eta'}^2 - \frac{1}{2} t}{
   \sqrt{t} } \ln \frac{ 2 m_{\eta'} + \sqrt{t} }{2 m_{\eta'} -
    \sqrt{t} } \Big) \no \\
&& - \frac{1}{8 \pi} \, m_\pi^2 \,
   \Big( \frac{1}{24 F_\pi^2} [D-3F]^2  \cos^2 \theta +
       \frac{1}{3 F_0^2} [D-3 \lambda_2]^2 \sin^2 \theta \no \\
&& + \frac{\sqrt{2}}{12 F_0 F_\pi} [D-3 \lambda_2] [D-3F]\sin 2 \theta \,\Big)
\, {\cal B} \, \Big( 2 m_\eta  +  \frac{ m_\eta^2 - \frac{1}{2} t}{
   \sqrt{t} } \ln \frac{ 2 m_\eta + \sqrt{t} }{2 m_\eta -
    \sqrt{t} }\Big) \no \\
&& + \frac{1}{8 \pi^2} \mnod \lambda_1 m_\pi^2 \, \Big( 
      {\cal A}  \cos^2 \theta \,\, 
      \Big[ - 1+ \ln \frac{m_{\eta'}^2}{\mu^2} 
           + 2 \sqrt{ \frac{ 4 m_{\eta'}^2}{t}
      - 1} \; \arcsin \frac{\sqrt{t} }{ 2 m_{\eta'}} \Big]\no \\
&& \qq \qq \qq \q + {\cal B} \sin^2 \theta \, \, 
      \Big[ - 1+  \ln \frac{m_\eta^2}{\mu^2} 
           + 2 \sqrt{ \frac{ 4 m_\eta^2}{t}
      - 1} \; \arcsin \frac{\sqrt{t} }{ 2 m_\eta}  \Big]  \Big) 
  + \Delta' \no \\ 
\eeqa
with the divergence
\beq
\Delta' =  4 \mnod \lambda_1 m_\pi^2  \, \Big( {\cal A}  \cos^2 \theta +
           {\cal B} \sin^2 \theta \Big) \, L .
\eeq

\newpage

%%%% figure captions

\section*{Figure captions}

\begin{enumerate}

\item[Fig.1] Shown is the self-energy diagram for the baryon octet.
         Solid and dashed lines denote baryons and
         pseudoscalar mesons, respectively.  

\item[Fig.2] Tadpole diagram contributing at leading order to the 
         baryon masses. 
         Solid and dashed lines denote the baryons and the meson
         singlet, respectively.  

\end{enumerate}

\newpage

%%%%%%% Figures   %%%%%%
\begin{center}
 
\begin{figure}[bth]
\centering
%\leavemode
\centerline{
\epsfbox{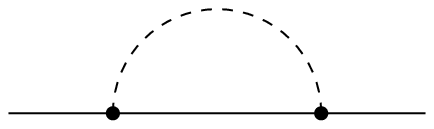}}
\end{figure}

\vskip 0.7cm

Figure 1

\vskip 1.5cm

\begin{figure}[tbh]
\centering
%\leavemode
\centerline{
\epsfbox{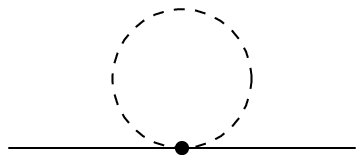}}
\end{figure}

\vskip 0.7cm

Figure 2

\end{center}

\end{document}